# Effect of Sr doping on the magnetic exchange interactions in manganites of type La$_{1-x}$Sr$_x$Mn$_y$A$_{1-y}$O$_3$ (A=Ga,Ti; 0.1≤y≤1)

A. Furrer[1,2], A. Podlesnyak[3], E. Pomjakushina[4], and V. Pomjakushin[1]

[1]*Laboratory for Neutron Scattering and Imaging, Paul Scherrer Institut, CH-5232 Villigen PSI, Switzerland*

[2]*SwissNeutronics AG, Bruehlstrasse 28, CH-5313 Klingnau, Switzerland*

[3]*Quantum Condensed Matter Division, Oak Ridge National Laboratory, Oak Ridge, TN 37831-6473, USA*

[4]*Laboratory for Scientific Developments and Novel Materials, Paul Scherrer Institut, CH-5232 Villigen PSI, Switzerland*

Strontium doping transforms manganites of type La$_{1-x}$Sr$_x$MnO$_3$ from an insulating antiferromagnet (x=0) to a metallic ferromagnet (x>0.16) due to the induced charge carriers (holes). Neutron scattering experiments were employed to investigate the effect of Sr doping on a tailor-made compound of composition La$_{0.7}$Sr$_{0.3}$Mn$_{0.1}$Ti$_{0.3}$Ga$_{0.6}$O$_3$. By the simultaneous doping with Sr$^{2+}$ and Ti$^{4+}$ ions the compound remains in the insulating state, so that the magnetic interactions for large Sr doping can be studied in the absence of charge carriers. At T$_C$=215 K there is a first-order reconstructive phase transition from the trigonal R-3c structure to the orthorhombic Pnma structure via an intermediate virtual configuration described by the common monoclinic subgroup P2$_1$/c. The magnetic excitations associated with Mn$^{3+}$ dimers give evidence for two different nearest-neighbor ferromagnetic exchange interactions, in contrast to the undoped compound LaMn$_y$A$_{1-y}$O$_3$ where both ferromagnetic and antiferromagnetic interactions are present. The doping induced changes of the exchange coupling originates from different Mn-O-Mn bond angles determined by neutron diffraction. The large fourth-nearest-neighbor interaction found for metallic manganites is absent in the insulating state. We argue that the Ruderman-Kittel-Kasuya-Yosida interaction reasonably accounts for all the exchange couplings derived from the spin-wave dispersion in metallic manganites.

PACS numbers: 75.30.Et, 75.47.Lx, 78.70.Nx, 61.05.F-

## I. INTRODUCTION

Manganites of perovskite type and composition $R_{1-x}D_xMnO_3$ ($R^{3+}$ and $D^{2+}$ are rare-earth and alkaline-earth ions, respectively) have attracted great interest, since they exhibit colossal magnetoresistance [1] and multiferroic behavior [2]. The parent compounds $R^{3+}Mn^{3+}O_3^{2-}$ are insulating antiferromagnets [3], which turn into metallic ferromagnets upon sufficient hole doping through chemical substitution onto the $R^{3+}$ sites by $D^{2+}$ ions, thereby promoting part of the manganese ions into $Mn^{4+}$ [4]. The drastical doping-induced changes of the magnetic exchange interactions are usually explained by the Zener double-exchange model [5]. It is the aim of the present work to study the effect of Sr doping on the magnetic exchange interactions in the insulating state of a tailor-made compound exclusively containing $Mn^{3+}$ ions. Our main conclusion is that ferromagnetism also occurs in the absence of charge carriers (holes). Furthermore, we clarify some of the controversial issues concerning the magnetic exchange couplings which arose from spin-wave measurements performed for a series of $R_{1-x}D_xMnO_3$ compounds [6-9].

The situation is very clear for the insulating parent compound $LaMnO_3$ in which the oxygen octahedra around the $Mn^{3+}$ ions are considerably distorted due to the cooperative Jahn-Teller effect. An orbital ordering is established resulting in an A-type magnetic structure [3], in which the $Mn^{3+}$ spins have parallel and antiparallel orientations in the basal plane and perpendicular to this plane, respectively. The nearest-neighbor antiferromagnetic interlayer and ferromagnetic intralayer couplings $J_{1,1}$ and $J_{1,2}$ mediated by the $O_1$ and $O_2$ ions, respectively, were quantified by spin-wave experiments [10,11]. By doping with $D^{2+}$ ions, the compound becomes a ferromagnetic insulator for $0.05<x<0.10$, and finally turns into a ferromagnetic metal above doping levels of $0.15<x<0.20$. With increasing doping, the structural distortions become increasingly suppressed, since $Mn^{4+}$ is not a Jahn-Teller active ion.

As mentioned above, the magnetic exchange interactions are drastically different between the insulating (x=0) and metallic (x>0.15-0.20) states of $R_{1-x}D_xMnO_3$ as evidenced by spin-wave experiments performed for $Pr_{0.63}Sr_{0.37}MnO_3$ [6], $Sm_{0.55}Sr_{0.45}MnO_3$ [7], $La_{0.75}Ca_{0.25}MnO_3$ [8], $Pr_{0.7}Ca_{0.3}MnO_3$ [8], $Pr_{0.55}(Ca_{0.85}Sr_{0.15})_{0.45}MnO_3$ [8], and $La_{1-x}(Ca_{1-y}Sr_y)MnO_3$ [9]. For comparison, Table I lists the relevant exchange parameters $J_i$ as visualized in Fig. 1. First, the antiferromagnetic interplane coupling $J_{1,1}$ turns into a ferromagnetic coupling. Second, the sizes of the nearest-neighbor exchange parameters $J_{1,1}$ and $J_{1,2}$ become considerably enhanced. Third, the spin-wave dispersion exhibits a broadening and a softening near the boundaries of the Brillouin zone. Several explanations were given in the literature to explain all these features. Both the ferromagnetic ground state and the enhancement of the nearest-neighbor exchange parameters, whose origin is superexchange in the undoped compound, were attributed to the Zener double-exchange interaction. The broadening of the spin waves near the zone boundary was explained either by spin-wave damping within a Stoner model [6] or by a random spatial distribution of different nearest-neighbor exchange parameters $J_1$ resulting from defects in the mean magnetic structure [9]. The softening of the spin waves near the zone boundary was phenomenologically described by introducing a rather large fourth-nearest-neighbor exchange parameter $J_4=(0.2-0.6)J_1$ [6-9]. The large value of $J_4$ was suggested to be due to $(3z^2-r^2)$-type orbital correlations in the exchange path [7], but this model fails for the case of $La_{1-x}Sr_xMnO_3$ (see Fig. 4 in Ref. [7]). Another explanation was based on a strong coupling of the spin waves with optic phonons which intersect each other towards the zone boundary [12].

In order to clarify some of these controversial issues we performed neutron scattering experiments on a Sr doped manganite compound with tailor-made composition $La_{0.7}Sr_{0.3}Mn_{0.1}Ti_{0.3}Ga_{0.6}O_3$. The chemical engineering of the present compound can be understood as follows. (i) In order to study Mn dimer excitations by neutron spectroscopy, the Mn content has to be reduced in the parent compound $LaMnO_3$. As already shown for $LaMn_{0.1}Ga_{0.9}O_3$ [13], the large substitution of $Mn^{3+}$ ions by $Ga^{3+}$ ions changes neither the insulating character of the compound nor the essential physics concerning the (ferromagnetic and antiferromagnetic) exchange parameters. (ii) The 30% substitution of $La^{3+}$ ions by $Sr^{2+}$ ions introduces charge carriers, thereby making the compound metallic with ferromagnetic exchange parameters. (iii) In order to turn the compound back into the insulating state, 30% of the $Ga^{3+}$ ions are substituted by $Ti^{4+}$ ions in order to avoid the creation of charge carriers by the dopant $Sr^{2+}$ ions. By this procedure, the total charge of the cations is kept unchanged at +6 per formula unit, *i.e.*, all Mn ions remain in the trivalent state as confirmed by the magnetic susceptibility data (see Sec. II.A), so that the exchange interactions can be studied in the absence of charge carriers (holes).

As already demonstrated by inelastic neutron scattering (INS) experiments performed for $LaMn_{0.1}Ga_{0.9}O_3$ [13], very precise exchange parameters can be obtained from the magnetic dimer splittings present in a magnetically diluted compound. We applied this method to the compound $La_{0.7}Sr_{0.3}Mn_{0.1}Ti_{0.3}Ga_{0.6}O_3$. The magnetic excitation spectra exhibit two different dimer splittings which we associate with ferromagnetically coupled nearest-neighbor $Mn^{3+}$ ions and exchange parameters $J_{1,1}$ and $J_{1,2}$ mediated by the $O_1$ and $O_2$ ions, respectively. Since charge carriers are absent in our sample, the ferromagnetic nature of $J_{1,1}$ results from superexchange, which can be qualitatively explained by the Mn-Mn bonding characteristics determined by neutron diffraction experiments. No evidence was found for the presence of $Mn^{3+}$ dimer splittings associated with the fourth-nearest-neighbor coupling $J_4$. The existence of $J_4$ in doped metallic manganites is obviously due to the charge carriers and most likely originates from the Rudermann-Kittel-Kasuya-Yosida (RKKY) interaction, which also contributes to the substantial enhancement of the parameters $J_{1,1}$ and $J_{1,2}$.

The present work is organized as follows. The sample synthesis and characterization as well as the experimental procedures by neutron scattering are described in Sec. II. Sec. III presents the crystal structures derived from neutron diffraction. Sec. IV summarizes some basics for magnetic dimer excitations, followed in Sec. V by the data and analysis of the neutron spectroscopic measurements. The results are discussed in the final Sec. VI.

## II. EXPERIMENTAL

### A. Sample Synthesis

The sample of $La_{0.7}Sr_{0.3}Mn_{0.1}Ti_{0.3}Ga_{0..6}O_3$ was synthesised by a solid state reaction using $La_2O_3$, $SrCO_3$, $Ga_2O_3$, $TiO_2$, and $MnO_2$ of a minimum purity of 99.99%. The respective amounts of starting reagents were mixed, milled and calcinated at temperatures between 1250 and 1500°C during 100 hours in air with intermediate grinding. The phase purity of the resulting black powder was confirmed by x-ray diffraction using a D8 Advance Bruker AXS diffractometer with Cu K$\alpha$ radiation. The results of magnetic susceptibility measurements performed with use of a Quantum Design Magnetic Properties Measurement System are shown in Fig. 2, together with data obtained for $LaMn_{0.1}Ga_{0.9}O_3$. Similar data for the latter compound were published by Blasco *et al.* [14]. The Curie-Weiss constant C determined for T>200 K results in a spin quantum number s=1.96 for the Sr-doped compound which is very close to s=2 expected for $Mn^{3+}$ ions, *i.e.*, the trivalent state of the Mn ions is not affected by the ion substitutions. The slight downward turn of the data below T=200 K is due to the structural phase transition at $T_C$=215 K discussed in Sec. III. In addition, the deviations from the Curie-Weiss law below T=200 K show the expected superparamagnetic behavior due to the presence of $Mn^{3+}$ N-mers (N=2,3,4,...).

## B. Neutron Diffraction

The neutron powder diffraction experiments were carried out at the spallation neutron source SINQ at PSI Villigen using the high-resolution diffractometer for thermal neutrons HRPT [15] ($\lambda$=1.155 Å, high resolution mode with $\delta d/d=10^{-3}$) at temperatures between 2 and 300 K. The refinements of the crystal structures were carried out with the program FULLPROF [16], with use of its internal tables for scattering lengths.

## C. Neutron Spectroscopy

The INS experiments were carried out with use of the high-resolution time-of-flight spectrometer CNCS [17] at the spallation neutron source (SNS) at Oak Ridge National Laboratory. The sample was enclosed in an aluminum cylinder (12 mm diameter, 45 mm height) and placed into a He cryostat to achieve temperatures T≥1.7 K. Additional experiments were performed for the empty container as well as for vanadium to allow the correction of the raw data with respect to background, detector efficiency, absorption, and detailed balance according to standard procedures.

## III. CRYSTAL STRUCTURES FROM NEUTRON DIFFRACTION

At room temperature, the compound $La_{0.7}Sr_{0.3}Mn_{0.1}Ti_{0.3}Ga_{0.6}O_3$ has the rhombohedral space group R-3c with the structure parameters shown in Table II. On cooling below $T_C$=215 K there is a first-order phase transition to the orthorhombic crystal structure Pnma with the parameters also shown in Table II. Pnma and R-3c are not group-subgroup related, thus this type of reconstructive transformation involves an intermediate virtual configuration described by a common subgroup of the symmetry groups of the two end phases [18,19]. The transition is abrupt, *i.e.*, there is no conventional order parameter as for second-order phase transitions. For these types of reconstructive phase transitions, one usually considers that chemical bonds are broken and reconstructed again. With the help of the program COMSUBS [18,19] we found that the maximum common subgroup of R-3c and Pnma is P2$_1$/c (No.14) with the following transformations (from R-3c): F-point of the Brillouin Zone (BZ) **k**=[0,1/2,1], with basis transformation (-4/3,-2/3,1/3), (0,1,0), (-2/3,-1/3,-1/3), and origin shift (1/3,-5/6,1/6); (from Pnma): Γ-point of BZ, with basis transformation (0,1,0), (0,0,1), (1,0,0). The use of this subgroup P2$_1$/c allows the refinement of all data points above and below T$_C$ with the same type of crystal metric.

The results of such a fit are shown in Fig. 3. There is an abrupt change of all lattice constants as well as a jump of the unit cell volume around T$_C$. The phase transition is of first-oder type as manifested by the presence of a temperature hysteresis and the coexistence of high- and low-temperature structures in the interval 200-220 K upon warming. Fig. 4 shows the bond lengths B-O within a BO$_6$ octahedron and the B-O-B angles between neighboring BO$_6$ octahedra in La$_{0.7}$Sr$_{0.3}$Mn$_{0.1}$Ti$_{0.3}$Ga$_{0.6}$O$_3$ (we denote the formula as ABO$_3$ with A=La,Sr and B=Mn,Ti,Ga). In the high-symmetry R-3c phase the octahedra are regular with a single B-O bond length dictated by the symmetry. The transition to the orthorhombic space group Pnma allows three different bond lengths, but as one can see from Fig. 4, they remain the same within experimental error implying the absence of the cooperative JT-effect. The structure parameter that strongly changes at T$_C$ is the B-O-B angle between neighboring octahedra - the angle is split into two values that are different by 4 degrees.

The structures presented in Table II have been refined from diffraction data taken with significantly higher statistics than the data for the temperature scans presented in Fig. 4. The error bars listed in Table II are smaller than those shown in Fig. 4, but the values are consistent with each other. From the high-statistics data we have also refined the atomic occupancies $p_i$. To have non-divergent fits some occupancies must be fixed. The coherent scattering lengths for the nuclei in $La_{0.7}Sr_{0.3}Mn_{0.1}Ti_{0.3}Ga_{0..6}O_3$ are $b_{La}$=8.24, $b_{Sr}$=7.02, $b_{Mn}$=-3.73, $b_{Ti}$=-3.44, $b_{Ga}$=7.29, and $b_O$=5.8 fm. One can see that La and Sr, as well as Mn and Ti, are difficult to distinguish. In the data analysis with fixed occupancies $p_{La}$=0.7, $p_{Sr}$=0.3, and $p_{Mn}$=0.1 and assuming full occupancy of the B-site we found $p_{Ti}$=0.315(3) and $p_{Ga}$=0.585(3) at both T=2 K and T=300 K within error bars. For the oxygen we found $p_O$=2.97(1) and 2.95(2) at 2K and 300K, respectively, *i.e.*, the actual oxygen content from the diffraction data is very close to the stoichiometric value $p_O$=3.

The symmetry of the compounds $ABO_3$ depends on the tolerance factor

$$t = \frac{r_A + r_O}{(r_B + r_O)\sqrt{2}} \quad (1)$$

and the Jahn-Teller (JT) effect. $r_X$ denotes the ionic radius. In the present case, 90% of the ions at the B site are JT-inactive, favoring the high symmetry. For the calculation of the tolerance factor we used standard radii tabulated by Shannon [20] with coordination numbers 12, 6, and 2 for A, B, and O sites, respectively. We would like to note that sometimes non-standard coordinations are used in the literature, such as 6 for oxygen and 9 for the A cation, resulting in different values of t. In $ABO_3$ perovskites even without JT-active ions the crystal structure becomes less symmetric as the tolerance factor decreases. For example, in $Sr_{1-x}Ca_xMnO_3$ with non-JT ions $Mn^{4+}$, the crystal structure progressively changes from cubic to tetragonal and finally to orthorhombic with decreasing values of t [21]. In the presently studied compound $La_{0.7}Sr_{0.3}Mn_{0.1}Ti_{0.3}Ga_{0..6}O_3$ the tolerance factor is t=0.9856 for the rhombohedral R-3c structure at room temperature, whereas $LaMn_{0.1}Ga_{0.9}O_3$ has an orthorhombic Pnma structure at all temperatures T<300 K with t=0.9747 [13]. Apparently the above values of t are close to the border between R-3c and Pnma symmetries for this type of systems.

**IV. BASICS FOR MAGNETIC MULTIMER EXCITATIONS**

**A. Multimer formation**

Multimers of $Mn^{3+}$ ions in the compound $La_{0.7}Sr_{0.3}Mn_{0.1}Ti_{0.3}Ga_{0.6}O_3$ occur simply because of the random distribution of $Mn^{3+}$ ions over the sites of the pseudocubic perovskite lattice. In the present work we are primarily interested in the formation of $Mn^{3+}$ dimers, which occur with a statistical maximum relative to other multimers for the chosen 10% Mn content. Assuming a statistical distribution of the $Mn^{3+}$ ions, the multimer probabilities can easily be calculated by elementary probability theory: We find probabilities of 53.1%, 20.9%, and 8.2% for monomers, dimers, and trimers, respectively [22].

**B. Spin Hamiltonian for a dimer**

We base the analysis of the $Mn^{3+}$ dimer transitions on the spin Hamiltonian

$$H = -2J_1 \mathbf{s}_1 \cdot \mathbf{s}_2 + D\left[(s_1^z)^2 + (s_2^z)^2\right] \quad . \tag{2}$$

where $\mathbf{s_i}$ denotes the spin operator of the magnetic ions, J the bilinear exchange interaction, and D the axial single-ion anisotropy parameter. The diagonalization of Eq. (2) is based on the dimer states |S,M>, where $\mathbf{S}=\mathbf{s_1}+\mathbf{s_2}$ is the total spin and -S≤M≤S. For D=0 and identical magnetic ions ($s_1=s_2$) the eigenvalues of Eq. (2) are degenerate with respect to the quantum number M:

$$E(S) = -J[S(S+1) - 2s_i(s_i+1)] \tag{3}$$

For $Mn^{3+}$ ions with $s_i=2$, ferromagnetic (J>0) and antiferromagnetic (J<0) exchange give rise to a nonet (S=4) and a singlet (S=0) ground state, respectively, as illustrated in Fig. 5. A non-zero anisotropy term (D≠0) has the effect of splitting the spin states |S> into the substates |S,±M>. For D<0 the energetic ordering of the sublevels |S,±M> has to be reversed in Fig. 5.

## C. Neutron cross-section for dimer transitions

For spin dimers the neutron cross-section for a transition from the initial state |S> to the final state |S'> is defined by [23]

$$\frac{d^2\sigma}{d\Omega d\omega} = \frac{N}{Z}(\gamma r_0)^2 \frac{k'}{k} F^2(Q) \exp\{-2W(Q)\} \exp\left\{-\frac{E(S)}{k_B T}\right\}$$
$$\times \frac{4}{3}\left[1 - (-1)^{\Delta S}\frac{\sin(QR)}{QR}\right] |\langle S\|T(s_1)\|S'\rangle|^2 \delta\{\hbar\omega + E(S) - E(S')\} . \tag{4}$$

where N is the total number of spin dimers in the sample, Z the partition function, k and k' the wave numbers of the incoming and scattered neutrons, respectively, Q the modulus of the scattering vector **Q=k-k'**, F(Q) the magnetic form factor, exp{-2W(Q)} the Debye-Waller factor, R the distance between the two dimer spins, $|\langle S\|T(s_1)\|S'\rangle|$ the reduced transition matrix element defined in Ref. [23], and $\hbar\omega$ the energy transfer. The remaining symbols have their usual meaning. The transition matrix element carries essential information to derive the selection rules for spin dimers:

$$\Delta S = S - S' = 0, \pm 1 \; ; \; \Delta M = M - M' = 0, \pm 1 . \tag{5}$$

The transitions for ferromagnetically and antiferromagnetically coupled $Mn^{3+}$ dimers observed in Ref. [13] as well as in the present work are marked in Fig. 5 by the arrows A and B, respectively.

**D. Probabilities for dimer formation**

The probability $w_1$ to find a given $Mn^{3+}$ ion in a dimer coupled by the nearest-neighbor exchange interaction $J_1$ is given by

$$w_1 = 6x(1-x)^{10} . \tag{6}$$

The probabilities $w_n$ for the formation of $Mn^{3+}$ dimers coupled by the $n^{th}$ nearest-neighbor exchange interactions $J_n$ are considerably smaller than $w_1$. In the present work we are interested in the probability $w_4$ associated with $Mn^{3+}$ dimers coupled by the fourth-nearest-neighbor exchange interaction $J_4$. Since spin-wave measurements reported vanishing or extremely small parameters $J_2$ and $J_3$ [6-9], we do not have to distinguish whether $Mn^{3+}$ or $Ti^{4+}$ and $Ga^{3+}$ ions are present at the second- and third-nearest-neighbor positions, thus the probability $w_4$ turns out to be

$$w_4 = 9x(1-x)^{21} . \tag{7}$$

For the manganese concentration $x=0.1$ we find $w_1=0.209$ and $w_4=0.098$. These probabilities are sufficiently large to allow the detection of the corresponding $Mn^{3+}$ dimer transitions in INS experiments.

## V. RESULTS AND ANALYSIS OF MAGNETIC DIMER EXCITATIONS

Energy spectra of neutrons scattered from $La_{0.7}Sr_{0.3}Mn_{0.1}Ti_{0.3}Ga_{0.6}O_3$ are shown in Fig. 6, together with the data obtained for $LaMn_{0.1}Ga_{0.9}O_3$ [13]. We did not observe magnetic excitations for energy transfers >2 meV. There are marked differences between the two compounds. For the Sr-doped sample, the lines $B_i$ corresponding to the $|0> \rightarrow |1>$ transition of antiferromagnetically coupled $Mn^{3+}$ dimers are absent, and the transition A is split into two lines. According to the energy splitting pattern of Fig. 5, the transition with largest energy is the $|4> \rightarrow |3>$ transition associated with ferromagnetically coupled $Mn^{3+}$ dimers. We therefore attribute the lines $A_1$ and $A_2$ to this transition. Our interpretation is supported by the Q dependence of the peak intensities as shown in Fig. 7, which is in agreement with the prediction of the cross-section formula (4) for nearest-neighbor $Mn^{3+}$ dimers. The lines $A_1$ and $A_2$ refer to out-of-plane and in-plane dimers, respectively, whose bond angles are different by 4 degrees (see Fig. 4). The observed intensity ratio $A_1/A_2=1/2$ is in agreement with the structure, since there are two and four nearest-neighbor $Mn^{3+}$ ions located at out-of-plane and in-plane positions, respectively. In conclusion, the out-of-plane exchange coupling, being antiferromagnetic for the compound $LaMn_xGa_{1-x}O_3$, turns into a ferromagnetic coupling for the Sr-doped compound.

The lines $A_1$ and $A_2$ are actually a superposition of eleven individual transitions of type $|4,\pm M> \rightarrow |3,\pm M'>$ due to the single-ion anisotropy parameter D as illustrated in Fig. 5, which could not be resolved in the experiments. For the data analysis we kept the anisotropy parameter fixed at the value D=0.036 meV determined in Ref. [13], and we described each line $A_1$ and $A_2$ by eleven Gaussian functions with equal linewidth FWHM=30 µeV corresponding to the instrumental energy resolution. The spectral strengths of the eleven Gaussians were fixed at the calculated transition probabilities. The background was approximated by an exponential function. The least-squares fitting procedure

converged to the parameters $J_{1,1}$=0.175(6) meV and $J_{1,2}$=0.210(7) meV as listed in Table I. The resulting energy spectra described by lines in Fig. 6 are in good agreement with the experimental data. Upon raising the temperature from 1.7 K to 7.0 K, the lines $A_i$ experience a marginal shift to lower energy transfers as predicted by the model calculations.

If $Mn^{3+}$ dimers coupled by the fourth-nearest-neighbor exchange interaction $J_4$ were present in the Sr-doped compound, a transition would be expected at an energy transfer between 0.4 and 1.0 meV with intensity comparable to the lines $A_i$, however, the observed energy spectra displayed in Fig. 6 do not give evidence for magnetic lines in that energy range. The existence of a large parameter $J_4$ obviously cannot be induced by magnetoelastic interactions [12], but it requires the Sr-doped manganites to be in the metallic state.

In addition to $Mn^{3+}$ dimers, other $Mn^{3+}$ multimers (mainly monomers and trimers) are present in the investigated compound as outlined in Section IV.A, thus their effect on the observed energy spectra has to be considered. The spin Hamiltonian of $Mn^{3+}$ monomers is governed by the single-ion anisotropy and is obtained by setting J=0 in Eq. (2). The corresponding ground-state is split into three doublets with energies 0 (for M=0), D (for M=±1), and 4D (for M=±2), thus the allowed transitions have energies ≤0.1 meV which could not be resolved from the elastic line. The situation for $Mn^{3+}$ trimers is more involved, since three different types of trimers have to be considered, where the $Mn^{3+}$ ions are coupled by the exchange parameters $J_{1,1}/J_{1,1}$, $J_{1,1}/J_{1,2}$, and $J_{1,2}/J_{1,2}$ for both straight and angled trimers. The trimer states can be defined as $|S_{12},S>$ where $S_{12}=s_1+s_2$ and $S=s_1+s_2+s_3$. The trimer ground state is $|4,6>$. Transitions are allowed to the excited states $|3,5>$ (at energies from 0.70 to 0.84 meV) and $|4,5>$ (at energies from 2.1 to 2.5 meV), which produces a considerable line broadening. Then, similar to the case of the dimer transitions outlined above, an additional line broadening of the order of 0.3-0.4 meV results from the splitting of the trimer transitions $|S_{12},S\rangle \rightarrow |S_{12}',S'\rangle$ into individual transitions $|S_{12},S,M\rangle \rightarrow |S_{12}',S',M'\rangle$ due to the single-ion anisotropy parameter D. All these line broadening effects smear out each trimer transition over an energy range of many tenths of meV, so that the trimer transitions (occurring with a probability about three times smaller than the dimer probability) could not be resolved in the experiments, but merely contribute to some unresolved intensity above the background level. Accordingly, the presence of the $|4,6\rangle \rightarrow |4.5\rangle$ trimer transition manifests itself in Fig. 6 as some excess intensity (data points above the calculated line) for energy transfers >2 meV.

Finally, a possible interpretation of the lines $A_1$ and $A_2$ in terms of $Mn^{3+}$ trimer transitions has to be excluded, since the corresponding intensities are governed by a Q-dependence which is drastically different from $Mn^{3+}$ dimer transitions. For instance, for trimer transitions with $\Delta S_{12}=\pm 1$, the Q-dependence of the intensities is identical to the dashed line in Fig. 7 [22].

## VI. DISCUSSION AND CONCLUDING REMARKS

Based on the results of the present neutron scattering study of a Sr-doped carrier-free manganite compound, we now address some of the open questions raised in Section I. The exchange parameter $J_{1,1}$, being antiferromagnetic in the Sr-free compound, becomes ferromagnetic upon doping with $Sr^{2+}$ ions. Since charge carriers are absent in our sample, the origin of the ferromagnetic coupling has to be related to the Mn-Mn bonding characteristics shown in Fig. 4 and discussed in Section III. In order to get both ferromagnetic and antiferromagnetic couplings in $ABO_3$, like in the A-type structure of $LaMnO_3$ with pure $Mn^{3+}$ ions, one needs to have cooperative JT ordering. The antiferrodistortive ordering of $z^2$ orbitals of $Mn^{3+}$ ions in the basal plane is responsible for the ferromagnetic coupling, because the superexchange interaction between empty and filled $e_g$ orbitals of neighboring $Mn^{3+}$ ions is ferromagnetic. The interplane superexchange between half-filled $t_{2g}$ orbitals is antiferromagnetic. In the present case of $La_{0.7}Sr_{0.3}Mn_{0.1}Ti_{0.3}Ga_{0.6}O_3$ we have completely regular average octahedra without any degree ordering of B-O bonds. Apparently it is more favorable for a pair of $Mn^{3+}$ ions to have locally always a perpendicular orientation of $z^2$ orbitals resulting in a ferromagnetic exchange independently of the $Mn^{3+}$ pair orientation - along or perpendicular to the basal plane. For $LaMn_{0.1}Ga_{0.9}O_3$ the distortion of the $BO_6$ octahedra is substantial [13], and the formal calculation of the filling of the $z^2$ orbital is 45%. The calculation was done from the B-O bond lengths following Ref. [24], similar to the calculations performed in Ref. [25], where the filling of the $z^2$ orbitals was 100% in the orbitally ordered phase of $(La_{1-y}Pr_y)_{0.7}Ca_{0.3}MnO_3$ for y=0.7. This partial bond order seems to be responsible for the presence of both ferromagnetic and antiferromagnetic couplings in $LaMn_{0.1}Ga_{0.9}O_3$.

It is interesting to compare the phase transition observed for $La_{0.7}Sr_{0.3}Mn_{0.1}Ti_{0.3}Ga_{0.6}O_3$ at $T_C$=215 K with the JT-active case of $(La_{1-y}Pr_y)_{0.7}Ca_{0.3}MnO_3$ (y = 0.7) [25] and pure $LaMnO_3$ [26]. In the above compounds there is a transition as a function of temperature from the orbitally ordered phase to an orbitally disordered phase, which is metrically cubic in both cases, but there is no cubic group that would fit the peak intensities, so the structure remains orthorhombic. Oppositely, in the present case the structures are drastically different by symmetry. The high temperature phase is protected by the R-3c symmetry against the JT-effect for the $e_g$ orbital. The low temperature Pnma phase could be JT-distorted by its symmetry, but it remains orbitally disordered, suggesting that the JT-effect is not the main reason for the structure transformation.

Our experiments give evidence for the existence of two nearest-neighbor exchange parameters $J_{1,1}$ and $J_{1,2}$ which are different from each other by 17%. The superexchange interaction can be phenomenologically described by [27]

$$J(r,\Theta)=J'[1+a_1\cos(\Theta)+a_2\cos^2(\Theta)]\exp(-br) \tag{8}$$

where J' is an energy prefactor, $\Theta$ the B-O-B angle, and r the B-O distance shown in Fig. 4. Both $J_{1,1}$ and $J_{1,2}$ have equal bond distances r, but the bond angles $\Theta$ differ by 4.6 degrees, resulting in 2.5% and 5.0% changes of $\cos(\Theta)$ and $\cos^2(\Theta)$, respectively. Depending on the prefactors $a_i$ (which can take values up to 5), the observed difference of $J_{1,1}$ and $J_{1,2}$ does not appear to be unrealistic. So far the distinction between $J_{1,1}$ and $J_{1,2}$ was not observed in spin-wave experiments on related manganites [6-9], possibly due to limited energy resolution. It is likely that the observed broadening of the spin waves near the boundaries of the Brillouin zone results from the presence of two different nearest-neighbor exchange interactions which have the largest broadening effect near the zone boundary. Similar thoughts were expressed in Ref. [9].

As concluded in Section V, the rather large fourth-nearest-neighbor interaction $J_4$ resulting from spin-wave experiments in the metallic state [6-9] is induced by the charge carriers. Whereas in the insulating state the magnetic coupling is predominantly due to superexchange, an additional carrier-mediated interaction is present in the metallic state. This is usually the Zener double-exchange mechanism, but the RKKY model is particularly useful to describe the long-range interaction between localized spins and free charge carriers. The carrier concentration in $La_{0.7}Sr_{0.3}MnO_3$ was reported to be $1.8 \cdot 10^{22}$ cm$^{-3}$ [28]. Actually, within the molecular-field approximation, the double-exchange and the RKKY models are equivalent [29]. The RKKY interaction predicts an oscillating exchange parameter J(R) with distance R from the localized spin:

$$J(R)=\alpha F(x),\, F(x)=\frac{x\cos(x)-\sin(x)}{x^4},\, x=2k_F R\,, \tag{9}$$

where α is an energy prefactor and $k_F$ the Fermi wavenumber. Fig. 8 shows the oscillatory behavior of $F(2k_FR)$ for $k_F$=(1.02-1.06) Å$^{-1}$ which roughly matches the size of the cuboid-like Fermi surface determined for $La_{0.7}Sr_{0.3}MnO_3$ by electron-positron annihilation [30]. The overall features of $F(2k_FR)$ confirm the results of the spin-wave experiments performed for doped manganites [6-9]. $J(R_1)$ clearly produces an enhancement of the ferromagnetic exchange couplings $J_{1,1}$ and $J_{1,2}$ compared to the parent compound. $J(R_4)$ is also ferromagnetic and confirms the large value of $J_4$. $J(R_3)\approx 0$ is consistent with the vanishing parameter $J_3$. $J(R_2)$ turns out to be antiferromagnetic, but most likely the competition with the ferromagnetic next-nearest-neighbor superexchange interaction results in a vanishing parameter $J_2$.

In conclusion, the present neutron scattering experiments of the compound $La_{0.7}Sr_{0.3}Mn_{0.1}Ti_{0.3}Ga_{0.6}O_3$ allowed to study the exchange interaction in manganites for large Sr doping in the insulating state. By this procedure the effect of Sr doping due to the structural changes could be separated from the doping effect due the introduction of charge carriers. The detailed structural information obtained by neutron diffraction turned out to be essential to understand the absence of an antiferromagnetic exchange coupling in Sr- and Ti-doped $LaMn_{0.1}Ga_{0.9}O_3$.

**Acknowlegments**

Part of this work was performed at the Swiss Spallation Neutron Source (SINQ), Paul Scherrer Institut (PSI), Villigen, Switzerland. Research at Oak Ridge National Laboratory's Spallation Neutron Source was supported by the Scientific User Facilities Division, Office of Basic Energy Sciences, U.S. Department of Energy.

# References

[1]  E. Dagotto, T. Hotta, and A. Moreo, Physics Reports **344**, 1 (2001).

[2]  T. Kimura, T. Goto, H. Shintani, K. Ishizaka, T. Arima, and Y. Tokura, Nature **426**, 55 (2003).

[3]  E. O. Wollan and W. C. Koehler, Phys. Rev. **100**, 545 (1955).

[4]  G. H. Jonker and J. H. van Santen, Physica (Utrecht) **16**, 337 (1950).

[5]  C. Zener, Phys. Rev. **82**, 403 (1951).

[6]  H. Y. Hwang, P. Dai, S-W. Cheong, G. Aeppli, D. A. Tennant, and H. A. Mook, Phys. Rev. Lett. **80**, 1316 (1998).

[7]  Y. Endoh, H. Hirata, Y. Tomioka, Y. Tokura, N. Nagaosa, and T. Fujiwara, Phys. Rev. Lett. **94**, 017206 (2005).

[8]  F. Ye, P. Dai, J. A. Fernandez-Baca, Hao Sha, J. W. Lynn, H. Kawano-Furukawa, Y. Tomioka, Y. Tokura, and Jiandi Zhang, Phys. Rev. Lett. **96**, 047204 (2006).

[9]  F. Moussa, M. Hennion, P. Kober-Lehouelleur, D. Reznik, S. Petit, H. Moudden, A. Ivanov, Ya. M. Mukovskii, R. Privezentsev, and F. Albenque-Rullier, Phys. Rev. B **76**, 064403 (2006).

[10] K. Hirota, N. Kaneko, A. Nishizawa, and Y. Endoh, J. Phys. Soc. Jpn **65**, 3736 (1996).

[11] F. Moussa, M. Hennion, J. Rodriguez-Carvajal, H. Moudden, L. Pinsard, and A. Revcolevschi, Phys. Rev. B **54**, 15149 (1996).

[12] P. Dai, H. Y. Hwang, J. Zhang, J. A. Fernandez-Baca, S.-W. Cheong, C. Kloc, Y. Tomioka, and Y. Tokura, Phys. Rev. B **61**, 9553 (2000).

[13] A. Furrer, E. Pomjakushina, V. Pomjakushin, J. P. Embs, and Th. Strässle, Phys. Rev. B **83**, 174442 (2011).

[14] J. Blasco, J. Garcia, J. Campo, M. C. Sanchez, and G. Subias, Phys. Rev. B **66**, 174431 (2002).

[15] P. Fischer, G. Frey, M. Koch, M. Koennecke, V. Pomjakushin, J. Schefer, R. Thut, N. Schlumpf, R. Bürge, U. Greuter, S. Bondt, and E. Berruyer, Physica **276-278**, 146 (2000).

[16] J. Rodríguez-Carvajal, Physica B **192**, 55 (1993). Also available via the Internet at: www.ill.eu/sites/fullprof/.

[17] G. Ehlers, A. A. Podlesnyak, J. L. Niedziela, E. B. Iverson, and P. E. Sokol, Rev. Sci. Instrum. **82**, 085108 (2011).

[18] B. J. Campbell, H. T. Stokes, D. E. Tanner, and D. M. Hatch, J. Appl. Crystallogr. **39**, 607 (2006). Also available via the Internet at: http://iso.byu.edu/iso/isotropy.php, ISOTROPY Software Suite.

[19] H. T. Stokes and D. M. Hatch, Phys. Rev. B **65**, 144114 (2002).

[20] R. D. Shannon, Acta Cryst. A **32**, 751 (1976).

[21] B. Dabrowski, O. Chmaissem, J. Mais, S. Kolesnik, J. D. Jorgensen, and S. Short, J. Solid State Chem. **170**, 154 (2003).

[22] A. Furrer, A. Podlesnyak, and K. W. Krämer, Phys. Rev. B **92**, 104415 (2015).

[23] A. Furrer and O. Waldmann, Rev. Mod. Phys. **85**, 367 (2013).

[24] J. Kanamori, J. Appl. Phys. **31**, S14 (1960).

[25] V. Yu. Pomjakushin, D. V. Sheptyakov, K. Conder, E. V. Pomjakushina, and A. M. Balagurov, Phys. Rev. B **75**, 054410 (2007).

[26] J. Rodriguez-Carvajal, M. Hennion, F. Moussa, A. H. Moudden, L. Pinsard, and A. Revcolevschi, Phys. Rev. B **57**, R3189 (1998).

[27] H. Weihe and H. U. Güdel, J. Am. Chem. Soc. **119**, 6539 (1997).

[28] A. S. Hamid and T. Arima, phys. stat. sol. (b) **241**, 345 (2004).

[29] J. König, J. Schliemann, T. Jungwirth, and A. H. MacDonald, in *Electronic Structure and Magnetism of Complex Materials*, edited by D. J. Singh and D. A. Papaconstantopoulos (Springer, Berlin, 2002), p. 163.

[30] E. A. Livesay, R. N. West, S. B. Dugdale, G. Santi, and T. Jarlborg, J. Phys.: Condens. Matter **11**, L279 (1999).

TABLE I. Mn-Mn exchange interactions in manganites. The exchange parameters $J_i$ correspond to a Heisenberg Hamiltonian $H=-2\Sigma_{ij}J_{ij}\mathbf{S_i}\cdot\mathbf{S_j}$. The nearest-neighbor exchange parameters $J_{1,1}$ and $J_{1,2}$ refer to different exchange paths mediated by oxygen ions at positions $O_1$ and $O_2$, respectively. The second- and third-nearest neighbor exchange parameters $J_2$ and $J_2$ are negligibly small.

| Compound | Structure | T[K] | $J_{1,1}$ [meV] | $J_{1,2}$ [meV] | $J_4$ [meV] | Reference |
|---|---|---|---|---|---|---|
| $LaMnO_3$ | Pbnm | 10 | -0.302(14) | 0.418(6) | 0 | [10] |
|  | Pbnm | 20 | -0.29(2) | 0.42(3) | 0 | [11] |
| $LaMn_{0.1}Ga_{0.9}O_3$ | Pbnm | 1.5 | -0.285(5) | 0.210(4) | 0 | [13] |
| $La_{0.7}Sr_{0.3}MnO_3$ | R-3c | 31 | 0.90(5) | 0.90(5) | 0.22(2) | [9] |
| $La_{0.7}Sr_{0.3}Mn_{0.1}Ti_{0.3}Ga_{0.6}O_3$ | Pnma | 1.7 | 0.175(5) | 0.210(5) | 0 | present work |

TABLE II. Structure parameters (lattice parameters a,b,c; fractional atomic coordinates x,y,z; isotropic displacement factor B; reliability factors $R_n$ and $\chi^2$ defined in Ref. [16]) of $La_{0.7}Sr_{0.3}Mn_{0.1}Ti_{0.3}Ga_{0.6}O_3$ at T=2 K and T=300 K with Pnma and R-3c (hexagonal settings) space groups, respectively. Atom positions for Pnma: La,Sr at (4c), Mn,Ti,Ga at (4a), $O_1$ at (4a), $O_2$ at (8d). Atom positions for R-3c: La,Sr at (6a), Mn,Ti,Ga at (6b), $O_1$ at (18e).

|  |  | T=2 K | T=300 K |
|---|---|---|---|
| a [Å] |  | 5.4908(1) | 5.52688(14) |
| b [Å] |  | 7.76247(14) | 5.52688(14) |
| c [Å] |  | 5.5295(1) | 13.4223(3) |
| La,Sr | x | 0 | 0 |
|  | y | 0 | 0 |
|  | z | 0.25 | 0.25 |
|  | B [Å$^2$] | 0.33(1) | 0.597(14) |
| Mn,Ti,Ga | x | 0 | 0 |
|  | y | 0 | 0 |
|  | z | 0.5 | 0 |
|  | B [Å$^2$] | 0.00(2) | 0.27(3) |
| $O_1$ | x | 0.5032(5) | 0.45319(7) |
|  | y | 0.25 | 0 |
|  | z | 0.05834(15) | 0.25 |
|  | B [Å$^2$] | 0.44(1) | 0.857(9) |
| $O_2$ | x | 0.2601(2) |  |
|  | y | 0.53092(7) |  |
|  | z | 0.2594(2) |  |
|  | B [Å$^2$] | 0.56(1) |  |
| $R_p$ [%] |  | 6.63 | 8.53 |
| $R_{wp}$ [%] |  | 6.86 | 8.30 |
| $R_{exp}$ [%] |  | 4.52 | 6.04 |
| $\chi^2$ |  | 2.30 | 1.89 |

**Figure Captions**

FIG. 1 (Color online) Schematic crystal structure of LaMnO$_3$ with magnetic exchange couplings J$_i$ indicated. Only the manganese ions (large spheres) and the oxygen ions (small spheres) are shown.

FIG. 2 (Color online) Inverse magnetic susceptibility vs temperature for LaMn$_{0.1}$Ga$_{0.9}$O$_3$ (full circles) and La$_{0.7}$Sr$_{0.3}$Mn$_{0.1}$Ti$_{0.3}$Ga$_{0.6}$O$_3$ (open circles). The dashed lines result from a fit to the Curie-Weiss law for T>200 K. C denotes the Curie-Weiss constant.

FIG. 3. Crystal metric as a function of temperature in La$_{0.7}$Sr$_{0.3}$Mn$_{0.1}$Ti$_{0.3}$Ga$_{0.6}$O$_3$ refined in the monoclinic space subgroup P2$_1$/c of both Pnma and R-3c space groups. The lattice constants a, b, and c are shown by rhombs, circles and squares, respectively. The open symbols are the data taken on sample warming, the filled symbols represent the data taken on cooling in the vicinity of the transition to demonstrate the presence of hysteresis. The monoclinic angle β (not shown) also has a step-like behaviour at the transition from about 90 to 89.5 degrees on warming. The inset shows the unit cell volume near the phase transition temperature T$_C$=215 K.

FIG. 4. The bond lengths B-O (filled symbols) and bond angles B-O-B (open symbols) as a function of temperature in La$_{0.7}$Sr$_{0.3}$Mn$_{0.1}$Ti$_{0.3}$Ga$_{0.6}$O$_3$ (ABO$_3$) refined in the orthorhombic space group Pnma below T$_C$ and in the rhombohedral space group R-3c above T$_C$.

FIG. 5 (Color online) Schematic sketch of energy level splittings of magnetic dimers with $s_i=2$. The arrows mark the transitions displayed in Fig. 6. The energy splittings of the states $|S\rangle$ into the substates $|S,\pm M\rangle$ resulting from the single-ion anisotropy are enhanced for better visualization.

FIG. 6 (Color online) Energy spectra of neutrons scattered from $LaMn_{0.1}Ga_{0.9}O_3$ [13] and $La_{0.7}Sr_{0.3}Mn_{0.1}Ti_{0.3}Ga_{0.6}O_3$ (present work) at T=1.7 K (full circles) and T=7 K (open circles). The arrows mark the observed transitions. The lines are the result of least-squares fitting procedures explained in the text.

FIG. 7 (Color online) Q-dependence of the neutron cross-section associated with $\Delta S=\pm 1$ transitions of $Mn^{3+}$ dimers. The lines correspond to the Q-dependent terms of Eq. (3) with different Mn-Mn bond distances $R_n$. The circles denote the intensities of the transitions $A_i$ observed for $La_{0.7}Sr_{0.3}Mn_{0.1}Ti_{0.3}Ga_{0.6}O_3$ at T=1.7 K.

FIG. 8 (Color online) Oscillatory behavior of the RKKY interaction defined in Eq. (8) for $\alpha=1$ and $k_F=(1.02-1.06)$ Å$^{-1}$. $R_n$ denotes the n*th* nearest-neighbor Mn-Mn distance for manganites of composition $R_{1-x}A_xMnO_3$.

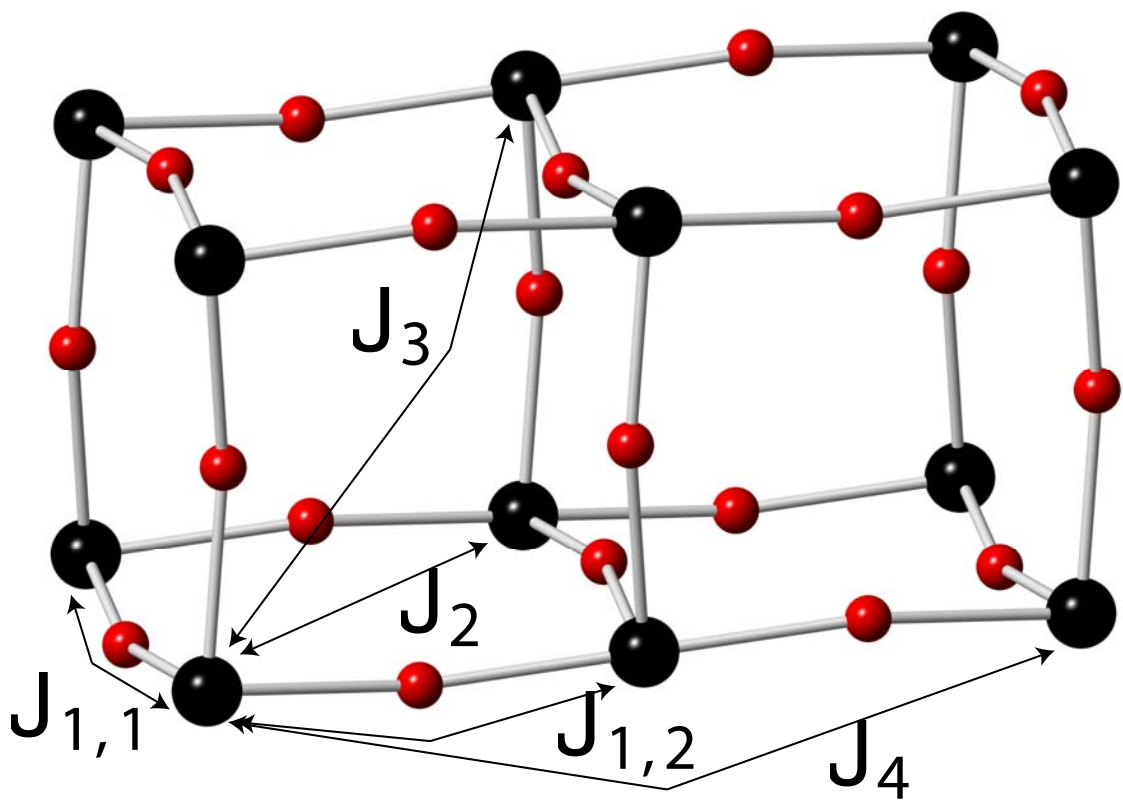

FIG. 1 (Color online) Schematic crystal structure of $LaMnO_3$ with magnetic exchange couplings $J_i$ indicated. Only the manganese ions (large spheres) and the oxygen ions (small spheres) are shown.

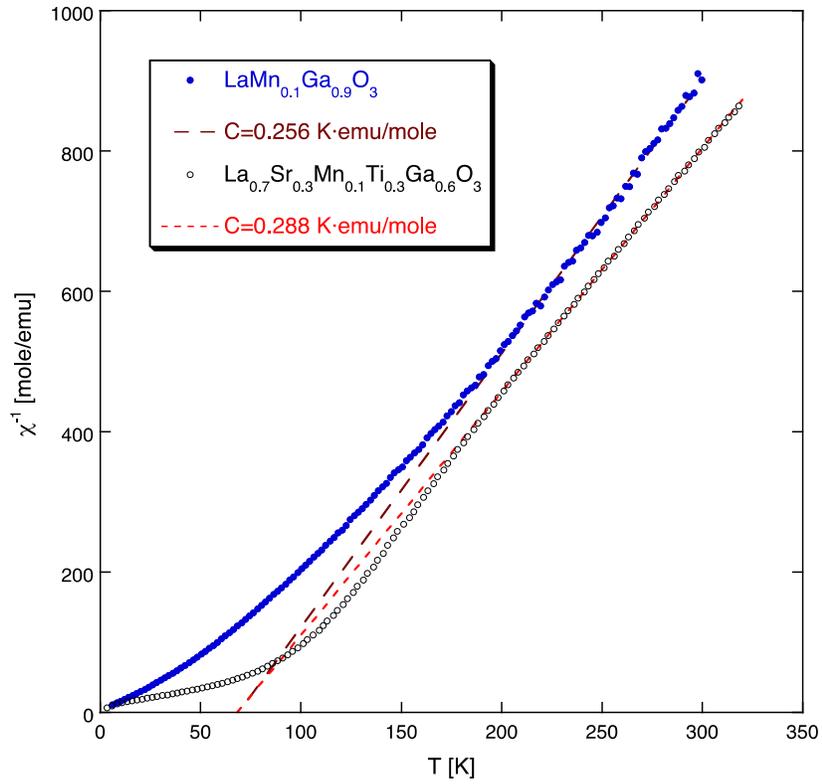

FIG. 2 (Color online) Inverse magnetic susceptibility vs temperature for LaMn$_{0.1}$Ga$_{0.9}$O$_3$ (full circles) and La$_{0.7}$Sr$_{0.3}$Mn$_{0.1}$Ti$_{0.3}$Ga$_{0.6}$O$_3$ (open circles). The dashed lines result from a fit to the Curie-Weiss law for T>200 K. C denotes the Curie-Weiss constant.

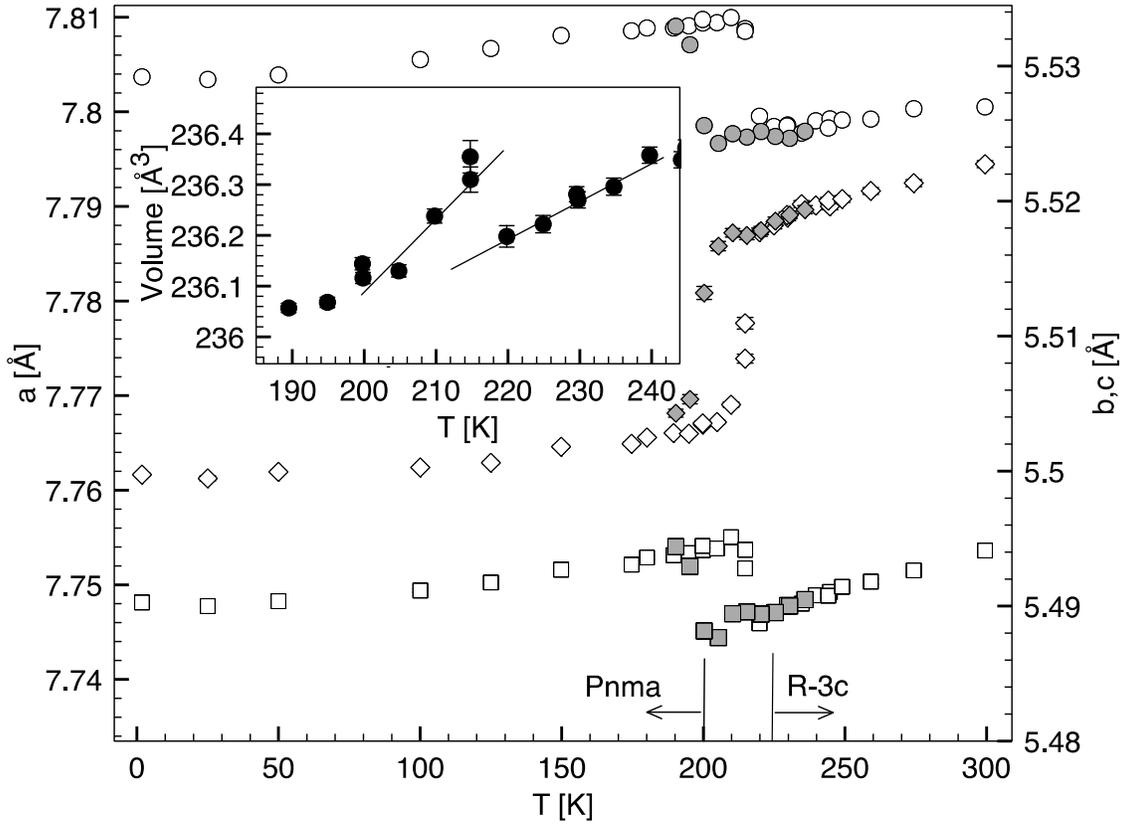

FIG. 3. Crystal metric as a function of temperature in $La_{0.7}Sr_{0.3}Mn_{0.1}Ti_{0.3}Ga_{0.6}O_3$ refined in the monoclinic space subgroup $P2_1/c$ of both Pnma and R-3c space groups. The lattice constants a, b, and c are shown by rhombs, circles and squares, respectively. The open symbols are the data taken on sample warming, the filled symbols represent the data taken on cooling in the vicinity of the transition to demonstrate the presence of hysteresis. The monoclinic angle β (not shown) also has a step-like behaviour at the transition from about 90 to 89.5 degrees on warming. The inset shows the unit cell volume near the phase transition temperature $T_C=215$ K.

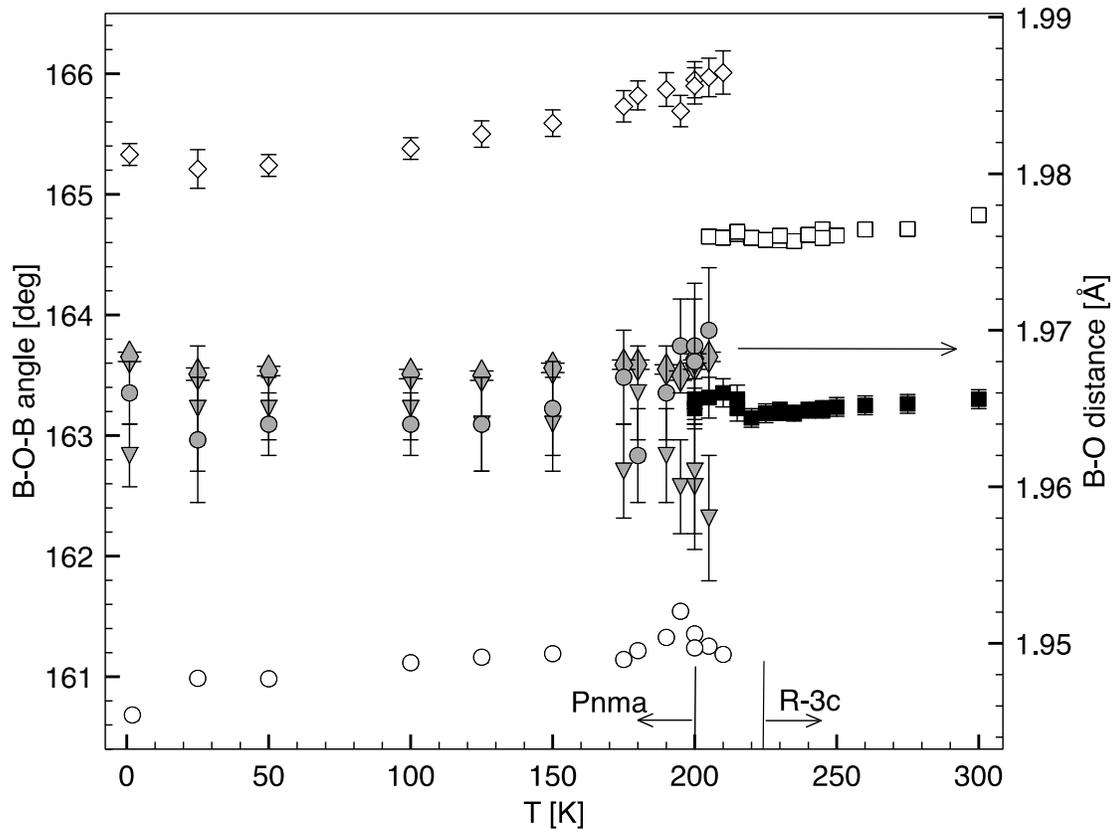

FIG. 4. The bond lengths B-O (filled symbols) and bond angles B-O-B (open symbols) as a function of temperature in $La_{0.7}Sr_{0.3}Mn_{0.1}Ti_{0.3}Ga_{0.6}O_3$ ($ABO_3$) refined in the orthorhombic space group Pnma below $T_C$ and in the rhombohedral space group R-3c above $T_C$.

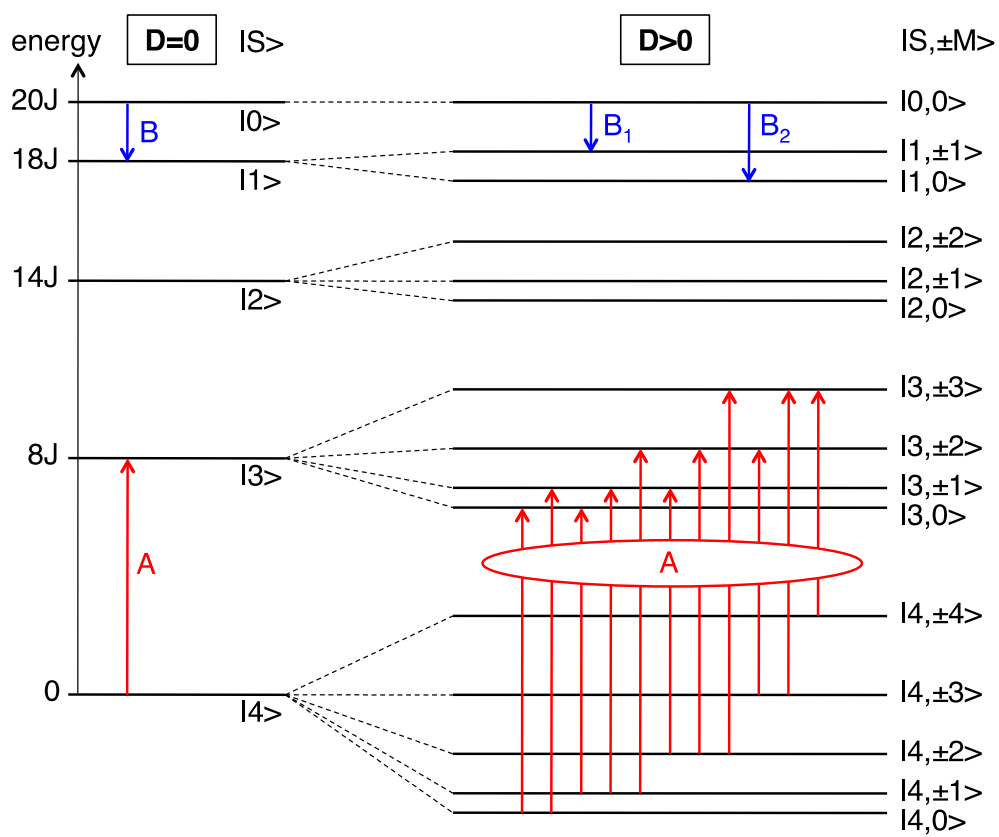

FIG. 5 (Color online) Schematic sketch of energy level splittings of magnetic dimers with $s_i=2$. The arrows mark the transitions displayed in Fig. 6. The energy splittings of the states $|S\rangle$ into the substates $|S,\pm M\rangle$ resulting from the single-ion anisotropy are enhanced for better visualization.

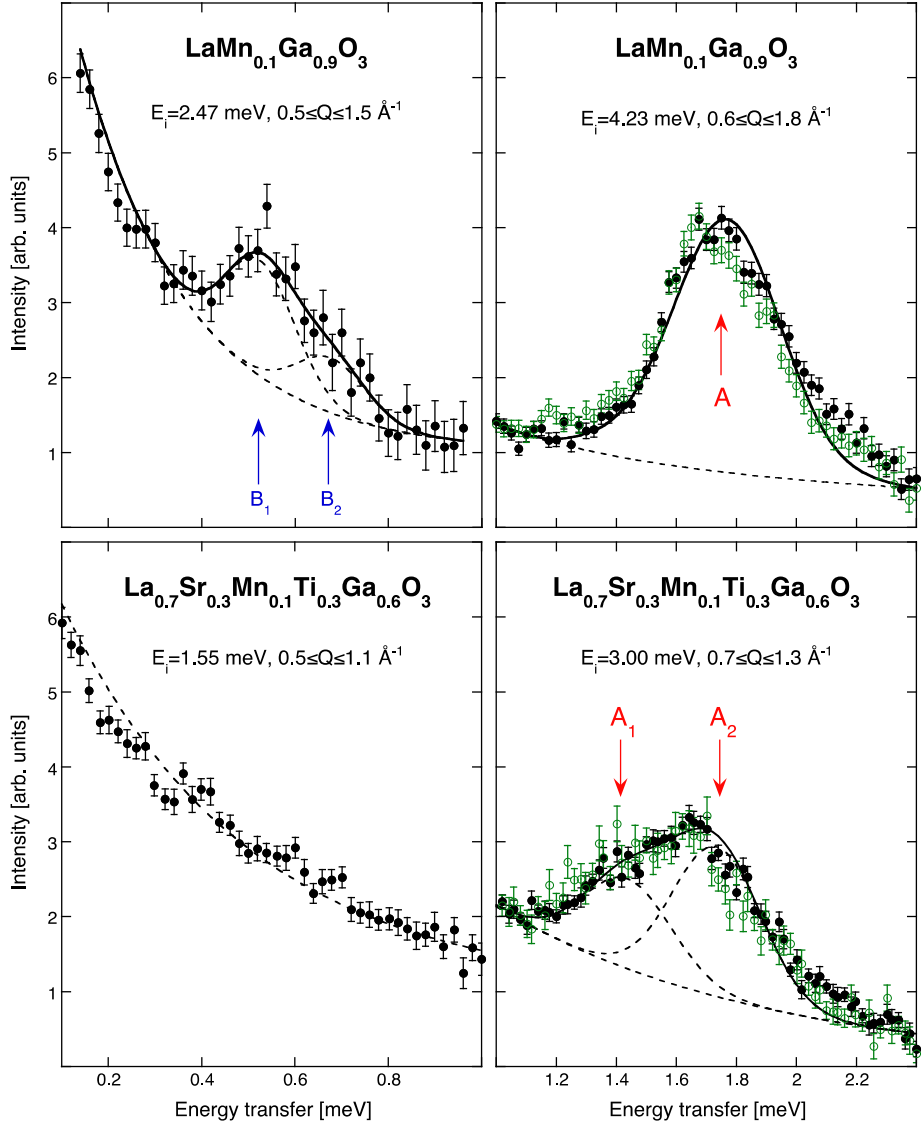

FIG. 6 (Color online) Energy spectra of neutrons scattered from LaMn$_{0.1}$Ga$_{0.9}$O$_3$ [13] and La$_{0.7}$Sr$_{0.3}$Mn$_{0.1}$Ti$_{0.3}$Ga$_{0.6}$O$_3$ (present work) at T=1.7 K (full circles) and T=7 K (open circles). The arrows mark the observed transitions. The lines are the result of least-squares fitting procedures explained in the text.

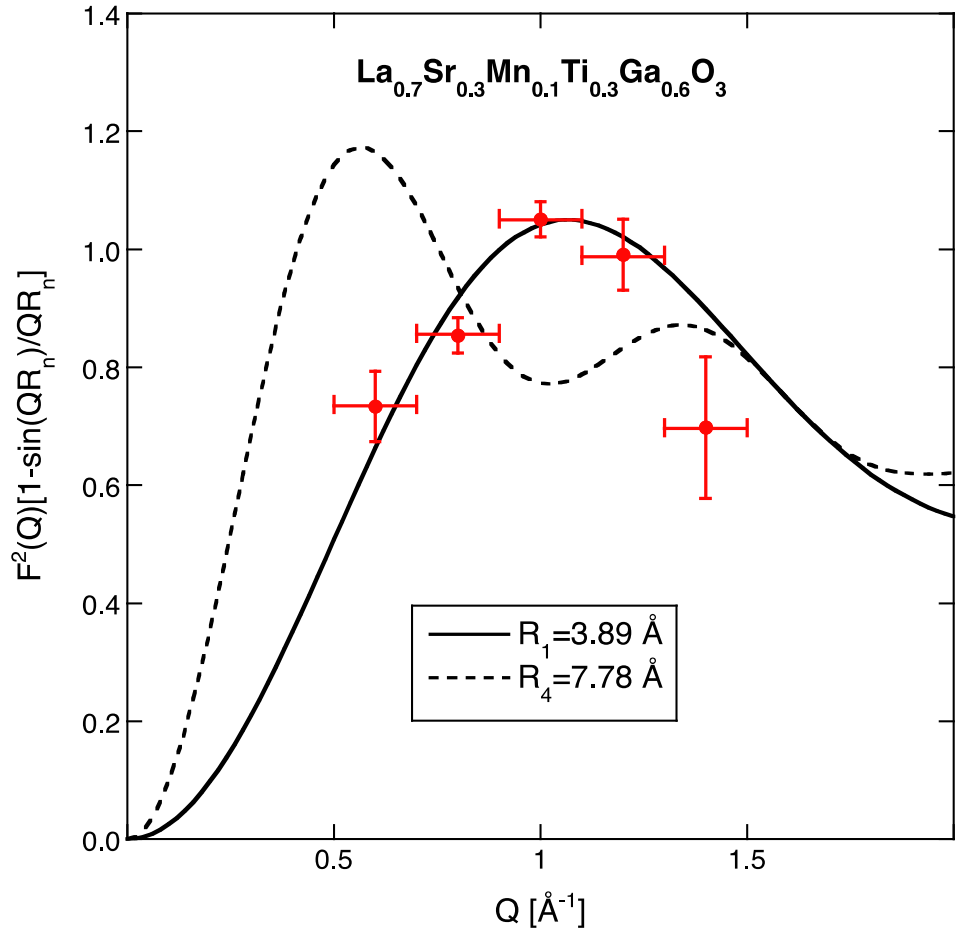

FIG. 7 (Color online) Q-dependence of the neutron cross-section associated with $\Delta S=\pm 1$ transitions of $Mn^{3+}$ dimers. The lines correspond to the Q-dependent terms of Eq. (3) with different Mn-Mn bond distances $R_n$. The circles denote the intensities of the transitions $A_i$ observed for $La_{0.7}Sr_{0.3}Mn_{0.1}Ti_{0.3}Ga_{0.6}O_3$ at T=1.7 K.

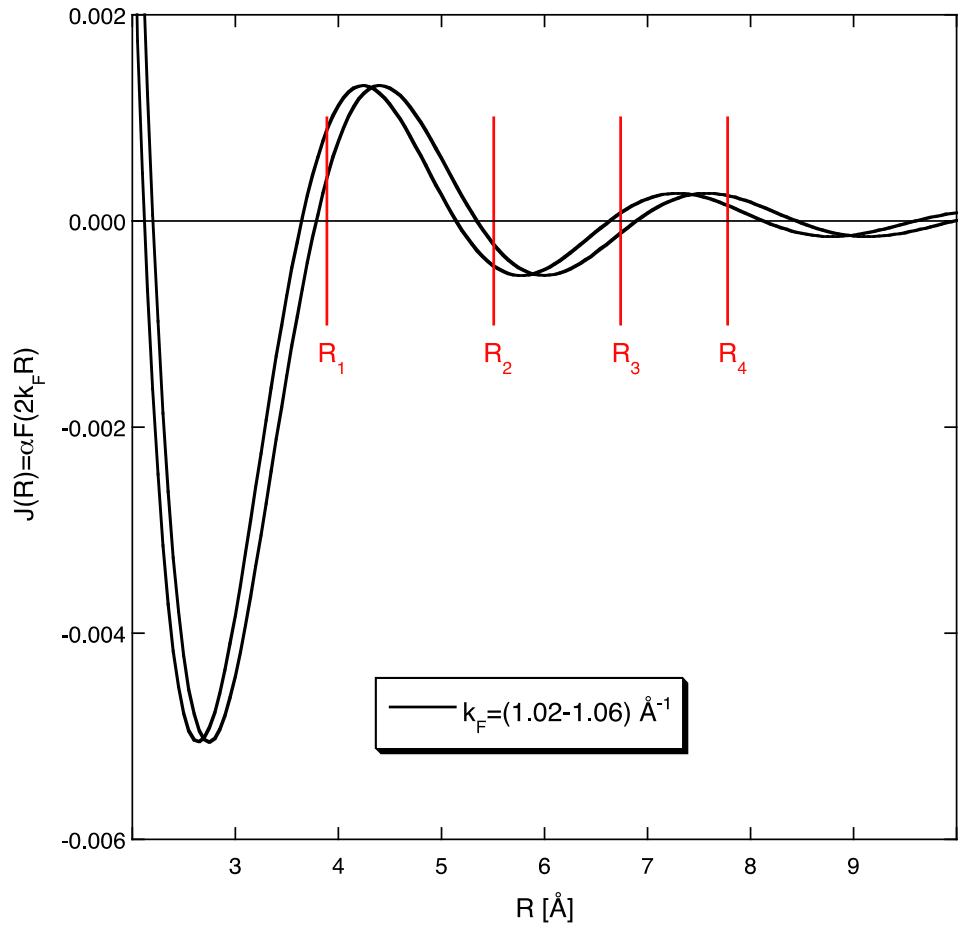

FIG. 8 (Color online) Oscillatory behavior of the RKKY interaction defined in Eq. (8) for $\alpha=1$ and $k_F=(1.02-1.06)$ Å$^{-1}$. $R_n$ denotes the n*th* nearest-neighbor Mn-Mn distance for manganites of composition $R_{1-x}A_xMnO_3$.